\documentclass[12pt]{iopart}

\usepackage{exscale}
\usepackage{iopams}
\usepackage{amsfonts}
\usepackage{enumerate}
\usepackage{graphicx}
\usepackage{cite}
\usepackage{times}

\usepackage{color}
\usepackage{soul}

\definecolor{darkred}{rgb}{0.9,0.1,0.1}
\definecolor{darkblue}{rgb}{0.1,0.1,0.7}

\usepackage{fancyhdr}
\newcommand{\bra}[1]{\langle{#1}\vert}
\newcommand{\ket}[1]{\vert{#1}\rangle}

\newcommand{\ee}{\mathrm{e}}
\newcommand{\ii}{\mathrm{i}}
\newcommand{\dd}{\mathrm{d}}

\newcommand{\an}[1]{{#1}^{\phantom\dag}}
\newcommand{\cre}[1]{{#1}^\dag}

\newcommand{\ev}[1]{\mathbb{E}\big[#1\big]}

\newcommand{\gen}{\mathcal{L}}
\newcommand{\para}{S}

\begin{document}

\vspace*{-50pt}

\title[Characteristic polynomial approach to inverse counting statistics]
{Inverse counting statistics for stochastic and open quantum systems: the characteristic polynomial approach}

\author{M~Bruderer$^1$, L~D~Contreras-Pulido$^1$, M~Thaller$^2$, L~Sironi$^3$, D~Obreschkow$^4$ and M~B~Plenio$^1$}

\address{$^1$Institut f\"{u}r Theoretische Physik, Albert-Einstein Allee 11, Universit\"{a}t Ulm, 89069 Ulm, Germany}

\address{$^2$Fachbereich Physik, Universit\"{a}t Konstanz, 78457 Konstanz, Germany}

\address{$^3$Department of Biology and Konstanz Research School Chemical Biology, University of Konstanz, 78457 Konstanz, Germany}

\address{$^4$The University of Western Australia, ICRAR, 35 Stirling Hwy, Crawley, WA 6009, Australia}

\ead{martin.bruderer@uni-ulm.de}

\date{\today}

\renewcommand{\baselinestretch}{0.95}
\begin{abstract}
We consider stochastic and open quantum systems with a finite number of states, where a stochastic
transition between two specific states is monitored by a detector. The long-time counting statistics
of the observed realizations of the transition, parametrized by cumulants, is the only
available information about the system. We present an analytical method for reconstructing generators 
of the time evolution of the system compatible with the observations. The practicality of the
reconstruction method is demonstrated by the examples of a laser-driven atom and the kinetics of
enzyme-catalyzed reactions. Moreover, we propose cumulant-based criteria for testing the non-classicality
and non-Markovianity of the time evolution, and lower bounds for the system dimension. Our analytical
results rely on the close connection between the cumulants of the counting statistics and the characteristic
polynomial of the generator, which takes the role of the cumulant generating function.
\end{abstract}
\renewcommand{\baselinestretch}{0.95}

\vspace{10pt}

\small
\tableofcontents
\normalsize

\maketitle



\section{Introduction}

When probing a physical system we often face the problem that only a small part
of its evolution is accessible to direct observation. Notably, in some cases,
the entire available information consists of observable time-discrete events,
indicating that a stochastic transition between two states of the system has been realized.
Examples of such events are photons emitted by fluorescent ions~\cite{Diedrich-PRL-1987}
or nitrogen-vacancy centers~\cite{Jelezko-PSS-2006}, single electrons passing through
quantum dots~\cite{Gustavsson-PRL-2006,Gustavsson-SSR-2009,Flindt-PNAS-2009},
individual steps of processive motor proteins~\cite{Kolomeisky-ARPC-2007} or product molecules
generated in enzyme-catalyzed reactions~\cite{English-NCB-2005,Moffitt-ME-2010,Moffitt-FEBS-2013}.
Common to all is that the observation of the system is restricted to a simple counting
process of certain discrete events by a detector.

In this paper we ask the question: What can be learned about the physical system
if the counting statistics of these events is the only available
information? To be specific, the long-time counting statistics is the discrete probability
distribution $p(n)$, expressing the probability that exactly $n$ events occur
within a sufficiently long time interval~$[0,t]$. The distribution $p(n)$ is
conveniently parametrized in terms of its cumulants, where non-zero higher order
cumulants indicate deviations from Gaussian behaviour. Our goal is to exploit
these deviations in order to recover the hidden structure and evolution of
the system from the observational data.

\begin{figure}[h]
\vspace*{15pt}
\centering
\includegraphics[width=0.75\columnwidth]{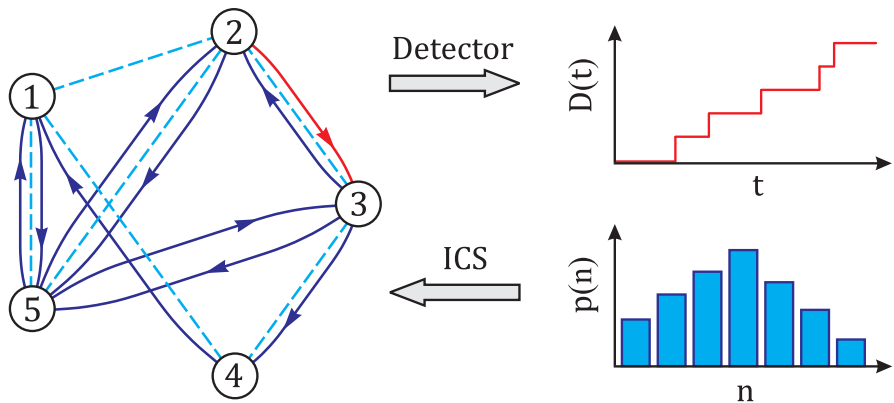}
\caption{A detector monitors a stochastic transition between two specific states (red arrow)
of a system with $N$ states connected by stochastic and quantum
transitions (arrows and dashed lines). The long-time counting statistics of the events
$p(n)$ obtained from the detector signal $D(t)$ contains sufficient information to
reconstruct the generators governing the time evolution of the system.}
\label{scheme}
\end{figure}

In answering the above question, we present several results that elucidate the
relation between the evolution of the system and the observed counting
statistics. Most importantly, we identify the characteristic polynomial of
the time-independent generator, governing the time evolution, as the central object
of our study. The characteristic polynomial is shown to have two essential
properties: First, it can be reconstructed from a finite number of cumulants, and second, 
it takes the role of the cumulant generating function. As an
immediate consequence, we find that for an open quantum system with dimension $N$
at most $2(N^2 - 1)$ cumulants of the observed counting statistics are independent.
Building on these results, we further design cumulant-based tests to identify
non-classical and non-Markovian time evolutions, and to estimate the
system dimension $N$. Moreover, we suggest an analytical method for reconstructing generators
compatible with the observations. We find that the result of the reconstruction process is not
necessarily unique, i.e.,~different generators may lead to identical observed counting statistics.
By analogy to problems such as inverse scattering theory~\cite{Kirsch-2011}
we refer to our methods collectively as \emph{inverse counting statistics} (ICS).

The theory of ICS is a new member of a family of methods to discern the properties
of a system from observational data.
Part of these methods have the objective of reconstructing specific system properties, such as quantum state
tomography~\cite{Vogel-PRA-1989,Paris-2004,Baumgratz-PRL-2013,Baumgratz-AX-2013}, which addresses
the problem of efficiently determining the \emph{state} of a physical system by measuring
a complete set of observables.
Various methods, on the other hand, are based on inequalities for selected system properties; these inequalities are
tested against experimental observations.
The most established test for the non-classicality of the \emph{evolution} of a
system is the Leggett-Garg inequality~\cite{Leggett-PRL-1985,Emary-AX-2013},
whose experimental violation was demonstrated with nitrogen-vacancy defects~\cite{Waldherr-PRL-2011}
and which may be used to evidence non-classical electron transport in
nano-structures~\cite{Lambert-PRL-2010}. The advantages of cumulant-based
inequalities for probing non-classicality of current correlations in mesoscopic junctions and
a related cumulant-based Bell test have been recently discussed in Refs.~\cite{Bednorz-PRL-2010,Bednorz-PRB-2011}.
Furthermore, identifying non-Markovian evolutions has become a subject of increasing
interest and several measures of non-Markovianity
in quantum systems have been proposed~\cite{Wolf-PRL-2008,Breuer-PRL-2009,Rivas-PRL-2010}.
For determining the dimension of quantum systems, the concept of dimension witness has
been introduced~\cite{Brunner-PRL-2008} and experimentally applied~\cite{Hendrych-NP-2012}.
The dimensionality of stochastic systems in the context of enzyme dynamics was shown to be
directly related to the so-called randomness parameter~\cite{Kolomeisky-ARPC-2007,Moffitt-ME-2010,Moffitt-FEBS-2013}.

The theory of ICS is of course related to the methods of full counting statistics
(FCS)~\cite{Levitov-JMP-1996,Belzig-PRL-2001,Bagrets-PRB-2003,Sanchez-PRL-2007,Flindt-PRL-2008,Flindt-PRB-2010,Nazarov-2009}
and large deviation theory~\cite{Derrida-PRL-1998,Lebowitz-JSP-1999,Derrida-JSM-2007,Touchette-PR-2009}.
These methods have been successfully applied to a wide range of stochastic and quantum systems, notably in mesoscopic
physics~\cite{Nazarov-2009}. In fact, measurements of higher order cumulants
of electronic current fluctuations have been achieved in tunnel junctions~\cite{Reulet-PRL-2003,Bomze-PRL-2005}
and in Coulomb blockaded quantum dots~\cite{Gustavsson-PRL-2006,Gustavsson-SSR-2009,Flindt-PNAS-2009}.
In general, FCS and large deviation theory are aimed at determining the counting statistics
of observable events if the generator of the time evolution and all system parameters are
known \emph{a priori}. Therefore these methods are not directly applicable to the inverse
problem considered here. 

The structure of this paper is as follows: In section~\ref{stwo} we introduce the underlying model and recall
the basics of the theory of FCS. In section~\ref{sthree} we give the details of the methods of ICS by establishing
the connection between the characteristic polynomial and the cumulants of the counting statistics. Furthermore,
we describe the cumulant-based tests and the analytical method for reconstructing the generator of the time
evolution. To illustrate the capacities of ICS we present specific reconstruction examples in section~\ref{sfour}. We end
with the conclusions in section~\ref{sfive}.


\section{Model and prerequisites}
\label{stwo}

While keeping the discussion broad we consider a specific system for clarity.
This system consists of $N$ orthogonal states, connected by stochastic and quantum
transitions (cf.~Fig~\ref{scheme}). The state of the system is described by the
density operator $\rho(t)$ whose time evolution is governed by the master equation
\begin{equation}\label{lindblad}
	\frac{\dd}{\dd t}\rho(t) = \gen\rho(t)\,,
\end{equation}
with $\gen$ the time-independent Markovian generator. The formal solution of Eq.~\eref{lindblad}
is given by $\rho(t)=\ee^{\gen t}\rho(0)$. We focus specifically on the long-time limit of the
evolution of the system and presume that Eq.~\eref{lindblad} has a unique steady state $\rho_s$
defined by $\gen\rho_s = 0$, i.e., the generator $\gen$ has a unique zero eigenvalue.

We want our model to cover two scenarios: In the classical limit, the system
evolves according to a time-continuous Markov process~\cite{Kampen-1992} with
only stochastic transitions, whereas in the general case we deal with an open
quantum system~\cite{Breuer-2002,Rivas-2011}. The model can also be seen as
a continuous-time extension of hidden quantum Markov models studied in Ref.~\cite{Monras-AX-2010}.
The generator $\gen$ in Lindblad form acts on $\rho$ as
\begin{equation}\label{qme}
    \gen\rho = -\ii[H,\rho] + \sum_{i\neq j}\Big(\an{L}_{ij}\rho\cre{L}_{ij} -
    \frac{1}{2}\{\cre{L}_{ij}\an{L}_{ij},\rho\}\Big)\,,
\end{equation}
where $\an{L}_{ij}$ and $\cre{L}_{ij}$ are Lindblad operators and $\{\:\,,\:\}$
stands here for the anticommutator. The Lindblad operators are of the form $L_{ij} =
\sqrt{\kappa_{ij}}\ket{i}\bra{j}$ and describe stochastic transitions from state
$\ket{j}$ to state $\ket{i}$ with time-independent rates $\kappa_{ij}\geq0$.
The time-independent Hamiltonian operator $H$ is expressed as a Hermitian matrix
in the basis $\{\ket{i}\}$, fixed by the stochastic transitions. \emph{A priori},
this model has $2N^2 - N$ real-valued parameters: $N^2 - N$ transition rates $\kappa_{ij}$,
$(N^2-N)/2$ complex-valued off-diagonal elements of $H$, and $N$ real-valued diagonal elements
of $H$. For many systems of interest, part of the stochastic or quantum transitions between
states are identically zero; the non-zero transitions are conveniently
depicted as a graph (cf.~Fig~\ref{scheme}).

We assume that the detector monitors a single stochastic transition $\ket{j_*}\rightarrow\ket{i_*}$
between two selected states $\ket{j_*}$ and $\ket{i_*}$ with perfect efficiency. The detector produces
a time-continuous signal $D(t)$, containing the number $n\geq 0$ of events observed up to the time $t$
(cf.~Fig.~1). The combined state of the system and detector after the $n$-th incoherent event is given
by the $n$-resolved density operator $\rho_n(t)$, where the additional index specifies the state of the
detector~\cite{Zoller-PRA-1987}.

The presented model is applicable to many interesting settings, in particular to stochastic
and quantum mechanical transport problems~\cite{Nazarov-2009,Kampen-1992}.
In this case, $N-1$ states describe different particle configurations and
the remaining state is identified with the empty state of the system.
Transitions that involve a change in the number of particles are interpreted as particle
transfers between the system and external reservoirs, and the detector measures
particle currents and fluctuations.


\subsection{Established results from FCS}

Let us briefly recall useful results from the theory of FCS (see Ref.~\cite{Flindt-PRB-2010} for details)
and define quantities of interest for the rest of the paper. The density operator 
of the system and detector $\rho_n(t)$ satisfies the $n$-resolved master equation
$\partial_t\rho_n(t) = \gen^{(0)}\rho_n(t) + \gen^{(1)}\rho_{n-1}(t)$,
where the generator has been separated into two parts as
$\gen = \gen^{(0)} + \gen^{(1)}$. In terms of matrix elements,
$\gen^{(1)} = \an{L}_{i_*j_*}\rho\cre{L}_{i_*j_*}$ contains the only
off-diagonal element corresponding to the transition $\ket{j_*}\rightarrow\ket{i_*}$
and $\gen^{(0)}$ the remaining elements. The finite-difference equation
for $\rho_n(t)$ is solved by using the discrete Laplace transform
$\rho_\xi(t) = \sum_n\rho_n(t)\ee^{\xi n}$ which obeys the equation
\begin{equation}\label{evol}
    \frac{\dd}{\dd t}\rho_\xi(t) = \gen_\xi\rho_\xi(t)\,,
\end{equation}
where the \emph{deformed generator} $\gen_\xi = \gen^{(0)} + \ee^{\xi}\gen^{(1)}$
has been introduced. The Fourier transform is often used instead of the Laplace transform,
which corresponds to the replacement $\xi\rightarrow\ii\xi$ throughout the paper.

A crucial observation of FCS and large deviation theory is that the cumulant generating function $G(\xi)=\log \ev{\ee^{\xi D}}$,
where $\ev{\!\cdot\!}$ stands for the expectation, is connected to the eigenvalue of smallest magnitude $\lambda(\xi)$
of the deformed generator $\gen_\xi$ through the relation $\lim_{t\rightarrow\infty}G(\xi)/t = \lambda(\xi)$~\cite{Derrida-PRL-1998,Touchette-PR-2009}.
In other words, $\lambda(\xi)$ corresponds to the unique zero eigenvalue of $\gen$ in the limit $\xi\rightarrow 0$.
In the long-time limit, the cumulants $C_\nu$ of the probability distribution $p(n)$ of the detector variable $D(t)$
are then given by
\begin{equation}\label{cumulants}
    C_\nu = \frac{\partial^\nu G(\xi)}{\partial \xi^\nu}\bigg|_{\xi=0}
    = t\,\frac{\partial^\nu \lambda(\xi)}{\partial \xi^\nu}\bigg|_{\xi=0}\qquad \nu\geq1\,.
\end{equation}
Using Eq.~\eref{cumulants} therefore allows us to determine the counting statistics of the detector variable $D(t)$,
in particular its average $\ev{D}=C_1$ and variance ${\rm Var}(D)=C_2$, provided that the deformed generator
$\gen_\xi$ is known. Since all cumulants $C_\nu$ increase linearly with time $t$, we introduce the scaled
cumulants $c_\nu = C_\nu/t$.

Corresponding results for the purely classical Markov process are obtained by restricting the previous
derivations to the occupation probabilities $\rho_{ii}$. It is then convenient
to introduce the probability vector $p = \{\rho_{ii}, i=1,\ldots,N\}$ and the stochastic generator $\gen^{\rm cl}$
acting on $p$. In practice, $\gen^{\rm cl}$ and the quantum generator $\gen^{\rm qm}$ are represented by $N\times N$
and $N^2\times N^2$ matrices, respectively, and $\rho = \{\rho_{ij},\,i,j=1,\ldots,N\}$ is a vector in Liouville
space with $N^2$ components. Unless necessary, we will not explicitly distinguish between stochastic and open quantum
systems and consider a generic generator of dimension $M\times M$.


\section{Inverse counting statistics}
\label{sthree}

We now present the general theory of ICS, which enables us to analyze the system by
using the information provided by measured cumulants. The main results of ICS are of two
kinds: The first (\S\ref{subsection_tests}) is formulated as a cumulant-based test
for non-classicality, non-Markovianity and system dimension $N$; the second
(\S\ref{subsection_generators}) aims at the reconstruction of generator of the time
evolution. Both are based on the close relation between the characteristic polynomial
of the generator and the cumulants, which will be established first.

\subsection{Characteristic polynomial of the generator}
\label{symmetry}

Most studies of stochastic or quantum systems focus on the spectrum of the generator,
e.g., the Hamiltonian of the system. The characteristic polynomial is rarely
used even though it contains the same information as the spectrum provided that the system
is finite-dimensional. A notable exception is the polynomial-based approach
to quantum mechanical perturbation theory by Raghunathan~\cite{Raghunathan-PIAS-1981}.
The significant advantage of the characteristic polynomial over the spectrum is that
it is always possible to obtain analytical expressions for the former.
This particular feature will be fully exploited in the following.
In contrast, analytical expressions for the spectrum, i.e., the roots of the characteristic polynomial,
cannot be found in general.

We here consider the characteristic polynomial $P_\xi(x) = \det[x\mathcal{I} -\gen_\xi]$ of the deformed
generator $\gen_\xi$, with $\mathcal{I}$ the identity matrix. Generally, $P_\xi(x)$ of degree $M$ can be
written as 
\begin{equation}\label{carpol1}
    P_{\xi}(x) = x^{M} + a_{M-1}x^{M-1} + \sum_{\mu=0}^{M-2}a_{\mu}(\xi)x^{\mu}\,.
\end{equation}
The coefficients $a_\mu(\xi)$ are given by the sum over the principal minors of order $(M-\mu)$ of
the deformed generator $\gen_\xi$~\cite{Macduffee-2004,Horn-2012} and depend on the variable $\xi$,
except for the coefficient $a_{M-1}$. The general expressions for the coefficients $a_\mu(\xi)$ are
unwieldy, but can be readily calculated for system dimensions of interest. Particularly simple exceptions
are $a_{M-1} = -\tr[\gen_\xi]$ and $a_0(\xi) = (-1)^M\det[\gen_\xi]$.  Note that each generator $\gen_\xi$
has a unique characteristic polynomial $P_\xi(x)$, whereas to a given generic polynomial there may correspond
none, one or several generators $\gen_\xi$.

In its factored form, $P_\xi(x)$  yields the entire spectrum $\sigma_\xi$ of the generator,
including $\lambda(\xi)$, and it thus contains more information than the cumulant generating
function $G(\xi)$. Consequently, generators $\gen_\xi$ with identical characteristic polynomials
$P_\xi(x)$ produce the same counting statistics and will be referred to as being equivalent.
The characteristic polynomial is therefore perfectly suited for studying symmetries with respect
to transformations of $\gen_\xi$ that leave the counting statistics unchanged. 
Specifically, $P_\xi(x)$ is invariant under arbitrary similarity transformations
of the generator $\gen_\xi\rightarrow T^{-1}\gen_\xi T$. However, the structure of $\gen_\xi$ imposed
by Eq.~\eref{qme} is not preserved under similarity transformations, i.e., the transformed matrix is
generally not a valid generator. Structure-preserving transformations are, for instance, permutations
applied to the matrix elements of $\gen_\xi$.  These transformations preserve the characteristic
polynomial $P_\xi(x)$ and the structure of $\gen_\xi$.

It is instructive to consider the example of a stochastic system with unidirectional transitions
between nearest-neighbours,~i.e.,~$\kappa_{i+1,i}>0$ for $i=1,\ldots, N$ with periodic boundary
conditions and all other rates zero. In this case, the generator $\gen_\xi^{\rm cl}$ is a triangular
matrix, except for a single element, and has the characteristic polynomial
\begin{equation}\label{sympol}
	P^{\rm cl}_\xi(x) = (-1)^N\left[\,\prod_{i=1}^N (x+\kappa_{i+1,i}) - \ee^\xi\prod_{i=1}^N \kappa_{i+1,i}\right].
\end{equation}
Clearly, $P^{\rm cl}_\xi(x)$ is invariant under any permutation of the rates $\kappa_{i+1,i}$ and hence
there are $N!$ equivalent generators related by permutations which yield identical counting statistics.
In addition, the statistics of the observations is independent of the position of the detector.


\subsection{Reconstruction of characteristic polynomials from cumulants}

The characteristic polynomial $P_\xi(x)$ plays an important role in ICS since it can be reconstructed
from a finite number of cumulants. Reconstructing the full generating function $G(\xi)$ from cumulants,
for example, by exploiting Eq.~\eref{cumulants}, may be possible for a few special cases. To our knowledge
there is however no general method available.

To start with, we parametrize $P_{\xi}(x)$ in a more convenient way. In view of the \emph{local}
dependence of the cumulants $c_\nu$ on the eigenvalue $\lambda(\xi)$ according to Eq.~\eref{cumulants}
we constrain our analysis to the properties of $P_{\xi}(x)$ in the vicinity of $\xi=0$. We 
consider the Taylor expansion of the coefficients $a_\mu(\xi)$ around $\xi=0$, which yields
\begin{equation}\label{taylor}
    P_{\xi}(x) = x^{M} + a_{M-1}x^{M-1} + \sum_{\mu=0}^{M-2}\sum_{k=0}^\infty
\frac{a_\mu^{(k)}\xi^k}{k!}\,x^{\mu}\,,
\end{equation}
with the shorthand notation $a_\mu^{(k)}\equiv\partial_\xi^k a_\mu(\xi)|_{\xi=0}$ and
in particular $a_\mu\equiv a_\mu(\xi)|_{\xi=0}$.
Since $a_\mu(\xi)$ depends on $\xi$ only through the factor $\ee^\xi$ in the generator
$\gen_\xi$ we find that $a_\mu^{(k)} = a_\mu^{(1)}\equiv a_\mu^\prime$ for all $k\geq 2$.
After simplifying Eq.~\eref{taylor} we can rewrite $P_\xi(x)$ in the compact form
\begin{equation}\label{taylorsimple}
    P_{\xi}(x) = P_\xi(x)\big|_{\xi=0} + \partial_\xi P_\xi(x)\big|_{\xi=0}\,(\ee^\xi-1)\,.
\end{equation}
The characteristic polynomial $P_\xi(x)$ in the vicinity of $\xi=0$ is thus fully determined by the
pair of polynomials $P_\xi(x)|_{\xi=0}$ and $\partial_\xi P_\xi(x)|_{\xi=0}$, or equivalently
parametrized by $2(M-1)$ of the coefficients $a_\mu$ and $a_\mu^\prime$, with $\mu=0,\ldots,M$. Explicitly,
the polynomials $P_\xi(x)\big|_{\xi=0}$ and $\partial_\xi P_\xi(x)\big|_{\xi=0}$ are specified
by the two sets of $M-1$ coefficients $\{a_1,\ldots,a_{M-1}\}$ and $\{a_0^\prime,\ldots,a_{M-2}^\prime\}$,
respectively, where we have taken into account that $a_0 = a_{M-1}^\prime = a_{M}^\prime = 0$ and
$a_M = 1$ are fixed.

Next, the direct relation between the cumulants $c_\nu$ and the
characteristic polynomial $P_\xi(x)$ is established. We note that the equality $P_{\xi}[\lambda(\xi)] = 0$
holds by definition of the characteristic polynomial. By repeatedly taking the total
derivative of this equality with respect to $\xi$ and evaluating it at $\xi = 0$, we can generate
the (infinite) set of equations
\begin{equation}\label{deriv}
    h_{\ell}(a_\mu,a_\mu^\prime,c_\nu)\equiv\frac{\dd^\ell P_{\xi}[\lambda(\xi)]}{\dd \xi^\ell}\bigg|_{\xi=0} = 0\qquad \ell\geq 1\,.
\end{equation}
We observe that Eqs.~\eref{deriv} relate the quantities $a_\mu$, $a^\prime_\mu$ and $c_\nu$, taking account of the relations
$c_\nu = \partial^{\nu}_{\xi}\lambda(\xi)|_{\xi=0}$ for $\nu\geq 1$ and $\lambda(\xi)|_{\xi=0} = 0$.
For instance, the first three functions $h_\ell(a_\mu,a_\mu^\prime,c_\nu)$ are given by
\begin{equation}\label{sample}
\eqalign{
    h_1 &= a_0^\prime + a_1 c_1\,, \\
    h_2 &= a_0^\prime + a_1 c_2 + 2a_1^\prime c_1 + 2a_2 c_1^2\,, \\
    h_3 &= a_0^\prime + a_1 c_3 + 3a_1^\prime (c_1 + c_2) + 6a_2 c_1 c_2 + 6a_2^\prime c_1^2 + 6a_3 c_1^3\,.
		}
\end{equation}
As can be seen from Eq.~\eref{taylorsimple} each function $h_\ell$ is linear in the coefficients
$a_\mu$, $a_\mu^\prime$. Moreover, considering $h_\ell$ as a function of $a_\mu$ and $a_\mu^\prime$,
we observe that only $h_\ell$ with $\ell\geq\mu$ depend on the coefficient $a_\mu$, and consequently
all $h_\ell$ with $\ell=1,\ldots,M$ are linearly independent, regardless of the values of the $c_\nu$.
Lastly, Eqs.~\eref{deriv} are inhomogeneous for $\ell\geq M$ because constant terms are generated by
the term $\lambda^M(\xi)$.

Using Eqs.~\eref{deriv} we can reconstruct the characteristic polynomial $P_\xi(x)$ from the first
$2(M-1)$ cumulants $c_\nu$. To this end, we choose
the first $2(M-1)$ equations $h_\ell(a_\mu,a_\mu^\prime,c_\nu) = 0$ with the cumulants $c_\nu$ as fixed
arguments and solve the resulting linear system for the coefficients $a_\mu$, $a_\mu^\prime$. Aside
from exceptional cases, the linear system has a unique solution for the $2(M-1)$ coefficients $a_\mu$
and $a_\mu^\prime$, thereby yielding a unique $P_\xi(x)$. We stress that the reconstruction of 
the characteristic polynomial $P_\xi(x)$ does not guarantee the existence of a generator $\gen_\xi$
compatible with the first $2(M-1)$ cumulants $c_\nu$. However, if the cumulants $c_\nu$ indeed result
from an evolution governed by a generator $\gen_\xi$ then the reconstructed $P_\xi(x)$ is the unique
characteristic polynomial of $\gen_\xi$.

From the reconstructed charateristic polynomial $P_\xi(x)$ all cumulants of higher order $c_\nu$, with
$\nu>2(M-1)$, can be found. One way of doing so is to evaluate $\lambda(\xi)$,
equivalent to the generating function $G(\xi)$, and subsequently use Eq.~\eref{cumulants}. We now introduce
a direct analytical method, which again is based solely on $P_\xi(x)$ and thus avoids the evaluation of $G(\xi)$.
Instead of solving Eqs.~\eref{deriv} for the coefficients $a_\mu$, $a_\mu^\prime$
with fixed cumulants $c_\nu$, we assume that the coefficients are known and solve Eqs.~\eref{deriv}
recursively for the cumulants $c_\nu$. We illustrate the procedure by evaluating the first three
(scaled) cumulants, the average, variance and skewness:
\begin{equation}\label{cm}
\eqalign{
c_1 &= -\frac{a_0^{\prime}}{a_1}\,, \\
c_2 &= -\frac{a_0^{\prime} + 2a_1^{\prime}c_1 + 2a_2c_1^2}{a_1}\,, \\
c_3 &= -\frac{a_0^{\prime} + 3a_1^{\prime}c_1 + 6a_2^{\prime}c_1^2 + 6a_3c_1^3 + 3a_1^{\prime}c_2 + 6a_2c_1c_2}{a_1}\,, \\
}
\end{equation}
where cumulants $c_\nu$ are expressed in terms of $a_\mu$, $a_\mu^\prime$ and cumulants $c_\eta$ with $\eta<\nu$.

In summary, we have two complementary methods at our disposal, which enable us (i) to reconstruct
$P_\xi(x)$ from a finite number of cumulants and (ii) to find all cumulants from a given $P_\xi(x)$. The
characteristic polynomial $P_\xi(x)$ thus completely replaces the cumulant generating function $G(\xi)$.
This remarkable property also suggests an alternative to the standard methods of FCS: Evaluating
(zero-frequency) cumulants within FCS usually requires the calculation of the eigenvalue $\lambda(\xi)$
or the regular part of the resolvant of $\gen_\xi$~\cite{Flindt-PRB-2010}. In this traditional way,
analytical results can only be obtained for small system dimensions. In contrast, our polynomial-based
procedure is efficient and direct: It suffices to calculate $P_\xi(x)$ for a given $\gen_\xi$, which yields
the coefficients $a_\mu\equiv a_\mu(\xi)|_{\xi=0}$ and $a_\mu^\prime\equiv\partial_\xi a_\mu(\xi)|_{\xi=0}$,
and then to use Eqs.~\eref{cm}. Analytical results for all cumulants $c_\nu$ can be obtained for any generator
$\gen_\xi$ regardless of the system dimension, including $\gen_\xi$ with more general Lindblad operators
than in Eq.~\eref{qme}. In particular, the Fano factor
\begin{equation}\label{fano}
	F\equiv \frac{c_2}{c_1} = 1 + \frac{2a_0^{\prime}a_2}{a_1^2} - \frac{2a_1^{\prime}}{a_1}
\end{equation}
is readily found in this way. In general, calculating cumulants recursively seems to
be more efficient than finding the full cumulant generating function $G(\xi)$. Recursive schemes for
calculating cumulants of high orders have also been exploited in traditional FCS~\cite{Sanchez-PRL-2007,Flindt-PRL-2008,Flindt-PRB-2010}.


\subsection{Tests for non-classicality, non-Markovianity and system dimension}
\label{subsection_tests}

\begin{table}[t]
\caption{We design several tests based on the fact that the number
of independent cumulants is at most $N_p$.  If more than $N_p$ independent
cumulants are observed in an experiment we can discard the hypothesis
under given prior assumptions.}
\label{tab}
\begin{center}
\begin{tabular}{l|l}
\emph{Prior Assumptions} & \emph{Hypothesis} \\ \hline
\\[-12pt]
Markovian, Dimension $N$ & Classicality \\
Classical, Dimension $N$ & Markovianity \\
Quantum, Dimension $N$ & Markovianity \\
Classical, Markovian & Dimension $N$ \\
Quantum, Markovian & Dimension $N$ \\
\end{tabular}
\end{center}
\end{table}

The reconstruction of the characteristic polynomial makes it possible to design tests for distinguishing
between different types of evolutions and system dimensions. The tests are based on the observation that
for a finite-dimensional Markovian system, with the Lindblad evolution according to Eq.~\eref{qme},
the number of independent cumulants is finite. Indeed, as shown previously, $P_\xi(x)$ can
be reconstructed from the first $N_{p} = 2(M-1)$ cumulants and serves as a cumulant generating function.
Thus, the number of independent cumulants is at most $N_{p}$. The specific value of $N_{p}$ depends on the
assumptions about the underlying model as regarding classicality, Markovianity and system dimension. If now one
of the assumptions is considered on the level of a hypothesis and the number of independent cumulants exceeds
the upper limit $N_{p}$, we can reject this initial hypothesis.

As a concrete example, consider a system with two dimensions and Markovian dynamics (`assumptions').
We want to test whether the dynamics is classical (`hypothesis') or not. The system may describe a
resonance fluorescence experiment with trapped ions~\cite{Diedrich-PRL-1987} or nitrogen-vacancy
centers~\cite{Jelezko-PSS-2006}, in which the counting statistics of emitted photons is analyzed.
Suppose the available cumulants are $\tilde{c}_1$, $\tilde{c}_2$ and $\tilde{c}_3$, where the tilde
marks quantities obtained from measurements. Given $\tilde{c}_1$ and $\tilde{c}_2$, and under the
hypothesis that the dynamics is classical we find from Eqs.~\eref{sample} for the classical case with $M=N$
\begin{equation}\label{pclass}
	\tilde{P}^{\rm cl}_\xi(x) = x^2 + \frac{2\tilde{c}_1^2}{\tilde{c}_1- \tilde{c}_2}x - \frac{2\tilde{c}_1^3}{\tilde{c}_1- \tilde{c}_2}(\ee^\xi-1)\,.
\end{equation}
The predicted classical value of the third cumulant $c^{\rm cl}_3 = \tilde{c}_1 + 3\tilde{c}_2(\tilde{c}_2/\tilde{c}_1-1)$
then follows from Eq.~\eref{cm}. If the predicted and measured value differ, i.e.,~$c^{\rm cl}_3\neq\tilde{c}_3$,
then the dynamics of the system is necessarily non-classical. This conclusion relies on the prior assumption
that the system is two-dimensional and Markovian, but the argument can be immediately adapted to different
combinations of assumptions and hypotheses, as summarized in Table~\ref{tab}. The generic structure of the
test is as follows:
\begin{enumerate}[(i)]
	\item The first $N_{p}+1$ cumulants of the counting statistics are measured.
	
	\item From the first $N_{p}$ cumulants the characteristic polynomial $P_\xi(x)$
	is reconstructed in accord with the prior assumptions (classicality, Markovianity
	or system dimension).
	
	\item The cumulant $c_{N_{p}+1}$ is predicted from $P_\xi(x)$
	and compared to the measured value of $c_{N_{p}+1}$.
	
	\item If the predicted and measured value of $c_{N_{p}+1}$ differ then the
	measured cumulant is independent from the lower-order cumulants.
	Therefore the dynamics of the system cannot be generated by $\gen_\xi$
	and the hypothesis is discarded.
\end{enumerate}
Inevitable experimental uncertainties lead of course to probabilistic rather than sharp
test results; however, the result can always be corroborated by measuring and comparing
more high-order cumulants and/or using a longer time base for the measurement.

Note that the test does not verify the existence of a generator $\gen_\xi$
compatible with the first $N_{p}$ cumulants; this problem is addressed in the next
section. As a consequence, the predicted and measured values of $c_{N_{p}+1}$ can
differ for two reasons: Either the generator $\gen_\xi$ compatible with the first
$N_{p}$ cumulants makes a false prediction about the measurement result,
or the generator does not exist, also leading to a false prediction.
In fact, the existence of a characteristic polynomial compatible with measurements
already imposes restrictions on the cumulants. We see from Eq.~\eref{pclass},
for instance, that for the classical two-state system it is required
that $\tilde{c}_1\neq\tilde{c}_2$. Moreover, as pointed out in Ref.~\cite{Brunner-PRL-2008},
one can only provide a lower bound on the unknown dimension of the system. In our case,
the test reveals that the system dimension must be at least $N_p+1$ and consequently we
cannot test for arbitrary dimensions~$N$.


\subsection{Reconstruction of generators from characteristic polynomials}
\label{subsection_generators}

Even though $P_\xi(x)$ contains essential information about the system, it is still desirable
to reconstruct the generator $\gen_\xi$ from cumulants. By using the generator together with
Eq.~\eref{lindblad} we can determine the full time evolution of the system and the steady
state $\rho_{s}$. The reconstruction furthermore allows us to find restrictions imposed on
the cumulants that warrant the existence of a generator.

For the reconstruction we have to solve the following inverse problem: Find the values
of the parameters entering the structured generator $\gen_\xi$ such that it reproduces
the observed cumulants $\tilde{c}_\nu$. The necessary and sufficient condition for $\gen_\xi$
to generate the cumulants $\tilde{c}_\nu$~is
\begin{equation}\label{poly0}
    P_\xi(x) = \tilde{P}_\xi(x)\,,
\end{equation}
where $\tilde{P}_\xi(x)$ is the reconstructed characteristic polynomial and $P_\xi(x)$
is the characteristic polynomial of $\gen_\xi$. We recall that while $P_\xi(x)$ is
invariant under similarity transformations, the structured generator $\gen_\xi$ is not,
which restricts the number of solutions to the inverse problem considerably.

Let us outline the direct analytical approach to solving the inverse problem. We exploit
the fact that analytical expressions are available for $P_\xi(x)$ and hence for the coefficients
$a_\mu$,~$a_\mu^\prime$. The latter are multivariate polynomials of degree $M-\mu$ in
the parameters of the model; the parameters are collected into the set $S$ for convenience.
Equation~\eref{poly0} is thus equivalent to the system of $2(M-1)$ polynomial equations
\begin{equation}\label{poly1}
\eqalign{
	a_\mu(\para) &= \tilde{a}_\mu\,, \\
	a_\mu^\prime(\para) &= \tilde{a}_\mu^\prime\,.
	}
\end{equation}
From a geometrical point of view, each equation defines an algebraic variety in the parameter
space corresponding to $\para$, and the solution of Eqs.~\eref{poly1} is given by the
intersection of these varieties~\cite{Cox-2007,Reilly-2010}.

It is not necessarily the case that all polynomial equations~\eref{poly1} are algebraically
independent, which can be checked, in principle, by using the standard methods of algebraic
geometry~\cite{Cox-2007,Reilly-2010}. To take this into account we introduce $N_i\leq 2(M-1)$,
the number of algebraically independent equations. If the number of parameters $|\para|$ is
larger than $N_i$, the generator $\gen_\xi$ cannot be determined by our approach. Otherwise,
if $|S|\leq N_i$, we can determine all structured $\gen_\xi$ compatible with Eq.~\eref{poly0}
in two steps: First, we select $|\para|$ independent equations from Eqs.~\eref{poly1} and solve
the resulting polynomial system, which reduces the original solution space to a finite discrete set,
denoted~$B$. Subsequently, we discard all solutions in $B$ that do not satisfy the full set of
Eqs.~\eref{poly1}, which finally leaves us with the solutions of the inverse problem.

The first step results in the polynomial system
\begin{equation}\label{poly2}
\eqalign{
    a_{\alpha}(\para) &= \tilde{a}_{\alpha}\,, \\
    a_{\beta}^\prime(\para) &= \tilde{a}_{\beta}^\prime\,,
		}
\end{equation}
where the indices $\alpha,\beta\in\{0,\ldots,M-1\}$ are used to select $|S|$ of the $N_i$
independent polynomial equations. In practice, it is advisable to select polynomials with
the smallest possible degrees for solving Eqs.~\eref{poly2}. At this point, we invoke
B\'{e}zout's theorem~\cite{Cox-2007,Reilly-2010}, stating that Eqs.~\eref{poly2} generally
have a \emph{finite} number of solutions, forming the set $B$. More precisely, if we admit
complex solutions, the number of distinct solutions is at most $\prod_k d_k$, where $d_k$ is the
degree of each polynomial in Eqs.~\eref{poly2}. This upper bound scales as $|\para|!$ and therefore
increases faster than exponentially with the system dimension $N$. The number of solutions can however be
significantly reduced by initially restricting the parameter space, most importantly,
to real-valued solutions for stochastic systems.

After selecting from $B$ all solutions that solve the full set of Eqs.~\eref{poly1} we are left
with zero, one or a finite number of solutions. No valid solution indicates that the underlying
model is not compatible with the observed cumulants $\tilde{c}_\nu$, e.g., false assumptions are
made about the dimension $N$. More than one solution is found if several equivalent generators
exist, which by definition cannot be distinguished by the detector. Additional information about
the system, such as state occupation probabilities, can be used to single out a unique generator
$\gen_\xi$. Alternatively, several independent detectors can be employed to reduce the number of
solutions.

\subsection{Embedding of the classical into a quantum model}

The dimension of $\gen_\xi$ is an important factor in the reconstruction because it sets the number
of conditions in Eqs.~\eref{poly1} and hence an upper bound for $N_i$. The number of conditions
scales as $N$ for classical systems and $N^2$ for quantum systems, whereas the number of parameters
$|S|$ typically scales as $N^2$ in both cases. Therefore, ICS applied to classical systems seems to
be limited to small dimensions because of the condition $|S|\leq N_i$.

We can avoid this problem to some extent by treating the classical system as quantum with a trivial Hamiltonian $H\equiv 0$
in Eq.~\eref{qme}, thereby embedding the classical into a quantum model. The resulting quantum generator $\gen^{\rm qm}_\xi$, according to
Eq.~\eref{qme}, for an actual classical system is sparse and block diagonal, i.e.,~$\gen^{\rm qm}_\xi$ separates into the
block $\gen^{\rm cl}_\xi$ and the diagonal block $\gen^{\rm coh}_\xi$ for the coherences, as exemplified by the generator
in Eq.~\eref{block}. The block $\gen^{\rm coh}_\xi$ contains elements stemming from the terms
$-\frac{1}{2}\{\cre{L}_{ij}\an{L}_{ij},\rho\}$ in Eq.~\eref{qme}, which in the long-time limit destroy
all coherences $\rho_{i\neq j}$ that may exist initially. As a consequence, the steady-state
evolutions resulting from the generators $\gen^{\rm qm}_\xi$ and $\gen^{\rm cl}_\xi$ are the same, and
the steady-state coherences are identically zero.

The point of interest for the reconstruction is of course the relation between the classical characteristic polynomial
$P^{\rm cl}_\xi(x)$ and quantum mechanical polynomial $P^{\rm qm}_\xi(x)$ obtained from the embedding. The characteristic
polynomial of the block-structured generator $\gen^{\rm qm}_\xi$ factorizes as
\begin{equation}
	P^{\rm qm}_\xi(x) = P^{\rm cl}_\xi(x) P^{\rm coh}_\xi(x)\,.
\end{equation}
The additional conditions we gain from treating the classical system as quantum mechanical therefore originate
from the polynomial $P^{\rm coh}_\xi(x)$, which can be made explicit by writing Eq.~\eref{poly0} in two parts
as $P^{\rm cl}_\xi(x) = \tilde{P}^{\rm cl}_\xi(x)$ and $P^{\rm coh}_\xi(x) = \tilde{P}^{\rm coh}_\xi(x)$.

In practice, starting from $2(N-1)$ measured cumulants of the classical system we reconstruct
$\tilde{P}^{\rm cl}_\xi(x)$, and in turn generate the first $2(N^2-1)$ cumulants. Using these
$2(N^2-1)$ cumulants we then proceed as for a quantum system to find the rates $\kappa_{ij}$.
In this way, the embedding strategy extends the use of ICS to classical systems of larger
dimension.


\section{Practical reconstruction examples}
\label{sfour}

\begin{figure}[t]
\centering
\includegraphics[width=0.35\columnwidth]{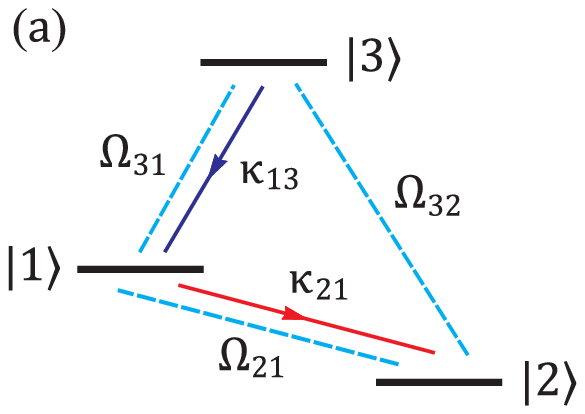}\hspace{20pt}\includegraphics[width=0.55\columnwidth]{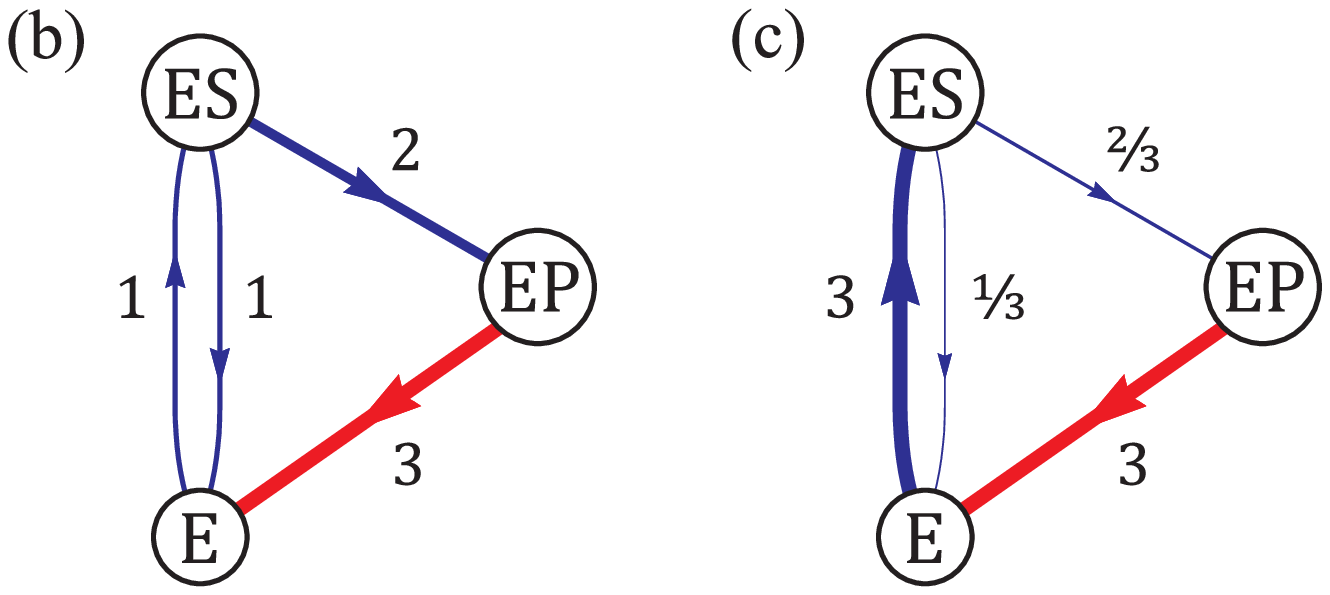}
\caption{The counting statistics of the monitored stochastic transition (red arrow) is used the reconstruct
the generator of the time evolution.
\textbf{(a)}~Three atomic states in $\Lambda$-type configuration are connected by stochastic and coherent transitions (arrows and dashed lines). The statistics
of the spontaneous decay of state $\ket{1}$ is used to determine the decay rates $\kappa_{21}$, $\kappa_{13}$ and Rabi frequencies
$\Omega_{32}$, $\Omega_{21}$, $\Omega_{31}$.
\textbf{(b)}~An enzyme-catalyzed reaction is modelled by stochastic transitions (arrows) between three states: empty enzyme $({\rm E})$,
enzyme-substrate complex $({\rm ES})$ and enzyme-product complex $({\rm EP})$. The counting statistics of the product molecules is used to
determine the transition rates. The observed statistics is identical for the two sets of rates shown in \textbf{(b)}~and~\textbf{(c)}.
}\label{bio}
\end{figure}

We illustrate the reconstruction of generators with concrete examples. For the stochastic
two-state system, we present the general solutions of Eqs.~\eref{poly2} in terms of the
measured cumulants $\tilde{c}_\nu$. As further examples we consider a laser-driven atomic
system and the Michaelis-Menten kinetics of enzymatic reactions. In the latter cases, we produce
the first $2(N^2-1)$ cumulants for a fixed set of parameters and afterwards reconstruct
generators $\gen_\xi$ compatible with these cumulants. Incidentally, this
application of ICS constitutes a general method for either verifying the uniqueness of a
generator or revealing symmetries of the system that are not immediately apparent from
the characteristic polynomial $P_\xi(x)$. An example for the characteristic polynomial
approach in connection with traditional FCS will be presented elsewhere~\cite{Contreras-TBP-2013}.


\subsection{Stochastic two-state system}

We first reconsider the classical two-state system with the detector at the transition
$\ket{1}\rightarrow\ket{2}$. The corresponding generator reads
\begin{equation}
	\gen_\xi^{\rm cl} = \bigg(
	\begin{array}{cc}
		-\kappa_{21} & \kappa_{12} \\
		\ee^\xi\kappa_{21} & -\kappa_{12}
	\end{array}\bigg)
\end{equation}
and the characteristic polynomial is
\begin{equation}
	P_\xi^{\rm cl} = x^2 + (\kappa_{21} + \kappa_{12})x - \kappa_{21}\kappa_{12}(\ee^\xi-1)\,.
\end{equation}
The reconstructed characteristic polynomial $\tilde{P}_\xi(x)$ is given by Eq.~\eref{pclass}.
To find the parameters $S=\{\kappa_{21},\kappa_{12}\}$ of the generator $\gen_\xi^{\rm cl}$
we have to solve Eq.~\eref{poly2}, i.e.,~the polynomial system
\begin{equation}
	 \tilde{a}_1 = \kappa_{21} + \kappa_{12} = \frac{2\tilde{c}_1^2}{\tilde{c}_1- \tilde{c}_2}\,, \qquad
	 \tilde{a}_0^\prime = -\kappa_{21}\kappa_{12} = -\frac{2\tilde{c}_1^3}{\tilde{c}_1- \tilde{c}_2}\,,
\end{equation}
with the solutions $\{\kappa_{21}, \kappa_{12}\} = \{\kappa_{+},\kappa_{-}\}$
and $\{\kappa_{21}, \kappa_{12}\} = \{\kappa_{-},\kappa_{+}\}$, where
\begin{equation}\label{twosols}
	\kappa_\pm = \frac{\tilde{c}_1^2\pm\tilde{c}_1^{3/2}\sqrt{2 \tilde{c}_2-\tilde{c}_1}}{\tilde{c}_1-\tilde{c}_2}\,.
\end{equation}
The fact that we obtain two solutions is in agreement with B\'{e}zout's theorem and reflected
in the symmetry of the characteristic polynomial $P_\xi^{\rm cl}$, which is a special
case of Eq.~\eref{sympol}. It follows from Eq.~\eref{twosols} that a classical generator
$\gen_\xi^{\rm cl}$ exists for cumulants restricted to the regime
$\tilde{c}_1>\tilde{c}_2\geq\frac{1}{2}\tilde{c}_1$. In particular, there is no generator
$\gen_\xi^{\rm cl}$ corresponding to an observed super-Poissonian ($\tilde{c}_2>\tilde{c}_1$)
counting statistics.


\subsection{Laser-driven atom with spontaneous decay}

We next consider a three atomic states in a $\Lambda$-type configuration~\cite{Fleischhauer-RMP-2005},
where the counting statistics is obtained by observing emitted photons (cf.~Fig.~\ref{bio}).
The three states $\ket{1}$, $\ket{2}$ and $\ket{3}$ are connected by coherent transitions,
parametrized by the Rabi frequencies $\Omega_{ij}$. We assume that the on-site energies of the
states are negligible, as is the case for near-resonant laser-driving, so that the
Hamiltonian $H$ has the form
\begin{equation}\label{ham}
    H = \frac{1}{2}\sum_{i>j}\Omega_{ij}\left(\ket{i}\bra{j} + \ket{j}\bra{i}\right)\,.
\end{equation}
Spontaneous decay is modelled by the stochastic transitions $\ket{3}\rightarrow\ket{1}$ and
$\ket{1}\rightarrow\ket{2}$ with decay rates $\kappa_{13}$ and $\kappa_{21}$, respectively,
where the latter is monitored by the detector. The generator of the evolution of the density matrix
$\rho = \{\rho_{11},\rho_{12},\ldots,\rho_{33}\}$ reads
\[\fl\gen_\xi^{\rm qm} =
\left(\hspace{-5pt}
\begin{array}{ccccccccc}
 -\kappa_{21} & \frac{\ii}{2}\Omega_{21} & \frac{\ii}{2}\Omega_{31} & -\frac{\ii}{2}\Omega_{21} & 0 &
0 & -\frac{\ii}{2}\Omega_{31} & 0 & \kappa_{13} \\
 \frac{\ii}{2}\Omega_{21} & -\frac{1}{2}\kappa_{21} & \frac{\ii}{2}\Omega_{32} & 0 & -\frac{\ii}{2}\Omega_{21} &
0 & 0 & -\frac{\ii}{2}\Omega_{31} & 0 \\
 \frac{\ii}{2}\Omega_{31} & \frac{\ii}{2}\Omega_{32} & -\frac{1}{2}\delta & 0 & 0 &
-\frac{\ii}{2}\Omega_{21} & 0 & 0 & -\frac{\ii}{2}\Omega_{31} \\
 -\frac{\ii}{2}\Omega_{21} & 0 & 0 & -\frac{1}{2}\kappa_{21} & \frac{\ii}{2}\Omega_{21} &
\frac{\ii}{2}\Omega_{31} & -\frac{\ii}{2}\Omega_{32} & 0 & 0 \\
 \ee^\xi\kappa_{21} & -\frac{\ii}{2}\Omega_{21} & 0 & \frac{\ii}{2}\Omega_{21} & 0 & \frac{\ii}{2}\Omega_{32} &
0 & -\frac{\ii}{2}\Omega_{32} & 0 \\
 0 & 0 & -\frac{\ii}{2}\Omega_{21} & \frac{\ii}{2}\Omega_{31} & \frac{\ii}{2}\Omega_{32} &
-\frac{1}{2}\kappa_{13} & 0 & 0 & -\frac{\ii}{2}\Omega_{32} \\
 -\frac{\ii}{2}\Omega_{31} & 0 & 0 & -\frac{\ii}{2}\Omega_{32} & 0 & 0 & -\frac{1}{2}\delta &
\frac{\ii}{2}\Omega_{21} & \frac{\ii}{2}\Omega_{31} \\
 0 & -\frac{\ii}{2}\Omega_{31} & 0 & 0 & -\frac{\ii}{2}\Omega_{32} & 0 & \frac{\ii}{2}\Omega_{21} &
-\frac{1}{2}\kappa_{13} & \frac{\ii}{2}\Omega_{32} \\
 0 & 0 & -\frac{\ii}{2}\Omega_{31} & 0 & 0 & -\frac{\ii}{2}\Omega_{32} & \frac{\ii}{2}\Omega_{31} &
\frac{\ii}{2}\Omega_{32} & -\kappa_{13} \\
\end{array}
\hspace{-3pt}\right),
\]
with $\delta = \kappa_{21} + \kappa_{13}$, from which the (albeit large) analytical expression for the
characteristic polynomial $P_\xi^{\rm qm}(x)$ is readily found.

For the set of dimensionless decay rates and Rabi frequencies $\{\kappa_{21}, \kappa_{13},\Omega_{32},\Omega_{21},
\Omega_{31}\}=\{5, 4, 3, 2, 1\}$ we produce the first $16$ cumulants $\tilde{c}_\nu = \{0.98,0.35,-0.04,\ldots\}$.
These are used to reconstruct the characteristic polynomial $\tilde{P}_\xi^{\rm qm}(x)$ in terms of the coefficients
$\tilde{a}_\mu$ and $\tilde{a}_\mu^\prime$. As a possible choice for the polynomial system in Eq.~\eref{poly2} we
select the five polynomial equations
\begin{equation}\label{atom}
\fl\qquad\eqalign{
\tilde{a}_6^\prime = -\frac{1}{2}\kappa_{21}\Omega_{21}^2\,, \\
\tilde{a}_5^\prime = -\frac{3}{2}\kappa_{13}\kappa_{21}\Omega_{21}^2 - \frac{1}{2}\kappa_{13}\kappa_{21}\Omega_{32}^2 - \frac{3}{4}\kappa_{21}^2\Omega_{21}^2\,, \\
\tilde{a}_8 = 3(\kappa_{13} + \kappa_{21})\,, \phantom{\frac{1}{1}}\\
\tilde{a}_7 = \frac{7}{2}(\kappa_{21}^2 + \kappa_{13}^2) + \frac{17}{2}\kappa_{13}\kappa_{21} + \frac{3}{2}(\Omega_{31}^2 + \Omega_{32}^2 + \Omega_{21}^2)\,, \\
\tilde{a}_6 = 4(\kappa_{21}\Omega_{32}^2 + \kappa_{13}\Omega_{21}^2) + \frac{37}{4}(\kappa_{21}\kappa_{13}^2 + \kappa_{13}\kappa_{21}^2)
 + \frac{13}{4}(\kappa_{21}\Omega_{31}^2 + \kappa_{13}\Omega_{32}^2) \\
						\qquad +\:2(\kappa_{13}^3 + \kappa_{21}^3) + \frac{11}{4}(\kappa_{21}\Omega_{21}^2 + \kappa_{13}\Omega_{31}^2)\,. \\
}
\end{equation}
The original parameters of the model are recovered by solving Eqs.~\eref{atom}, however, with all possible sign changes of the Rabi
frequencies, i.e., $\{\kappa_{21}, \kappa_{13},\Omega_{32},\Omega_{21},\Omega_{31}\}=\{5, 4, \pm 3, \pm 2, \pm 1\}$.
It can be immediately verified that the characteristic polynomial $P^{\rm qm}_\xi(x)$ and Eqs.~\eref{atom}
depend only on the square of the Rabi frequencies, which results in the sign symmetry with eight equivalent
generators. The example illustrates that by observing the spontaneous decay of state $\ket{1}$ we can determine
the magnitude of all Rabi frequencies $\Omega_{ij}$ and spontaneous decay rates $\kappa_{ij}$ of the atomic system.


\subsection{Michaelis-Menten kinetics with fluorescent product molecules}

Finally, we apply ICS to enzymatic reactions that are described by the Michaelis-Menten kinetics~\cite{English-NCB-2005,Moffitt-ME-2010,Moffitt-FEBS-2013}.
The kinetics of a single enzyme can be modelled by a stochastic three-state system, where the states correspond to empty enzyme
$({\rm E})\equiv\ket{1}$, the enzyme-substrate complex $({\rm ES})\equiv\ket{2}$ and the enzyme-product complex
$({\rm EP})\equiv\ket{3}$~\cite{Qian-BC-2002}. The enzyme-substrate binding is a reversible process while the other
transitions are assumed to be unidirectional (cf.~Fig.~\ref{bio}). The counting statistics is obtained from monitoring
the transition $({\rm EP})\rightarrow({\rm E})$ through the detection of single fluorescent product molecules~\cite{English-NCB-2005}.

For this biological scenario, we produce cumulants for the dimensionless rates $\{\kappa_{21},\kappa_{12},\kappa_{32},\kappa_{13}\}
=\{1, 1, 2, 3\}$ with the detector at the transition $\ket{3}\rightarrow\ket{1}$. Considering the system as being classical we obtain
the generator
\begin{equation}\gen_\xi^{\rm cl} = 
\left(
\begin{array}{ccc}
 -\kappa_{21} & \kappa_{12} & \ee^\xi\kappa_{13} \\
 \kappa_{21} & -\kappa_{12}-\kappa_{32} & 0 \\
 0 & \kappa_{32} & -\kappa_{13} \\
\end{array}
\right)
\end{equation}
and the characteristic polynomial
\begin{equation}\fl\qquad\eqalign{
	P_\xi^{\rm cl} &= x^3 + (\kappa_{12} + \kappa_{13} + \kappa_{21} + \kappa_{32})x^2 + (\kappa_{12}\kappa_{13}
	+ \kappa_{13}\kappa_{21} + \kappa_{13}\kappa_{32} + \kappa_{21}\kappa_{32})x \\
	&\qquad-\kappa_{13}\kappa_{21}\kappa_{32}(\ee^\xi-1)\,.
	}
\end{equation}
The classical polynomial $P_\xi^{\rm cl}(x)$ yields the three polynomial constraints 
\begin{equation}\label{mmclass}
\eqalign{
\tilde{a}_1 &= \kappa_{12}\kappa_{13} + \kappa_{13}\kappa_{21} + \kappa_{13}\kappa_{32}+ \kappa_{21}\kappa_{32}\,, \\
\tilde{a}_2 &= \kappa_{12} + \kappa_{13} +  \kappa_{21} +  \kappa_{32}\,, \\
\tilde{a}_0^\prime &= -\kappa_{13}\kappa_{21}\kappa_{32}\,,
}
\end{equation}
which are not sufficient to determine the four unknown rates $\kappa_{ij}$. Therefore,  we have to treat
the system as quantum mechanical and use $16$ cumulants to reconstruct the block-structured quantum generator
\begin{equation}\label{block}\fl
	\gen_\xi^{\rm qm} =
	\left(
\begin{array}{ccccccccc}
 -\kappa_{21} & \kappa_{12} & \ee^\xi\kappa_{13} &   &   &   &   &   &   \\
 \kappa_{21} & -\kappa_{12}-\kappa_{32} & 0 &   &   &   &   &   &   \\
 0 & \kappa_{32} & -\kappa_{13} &   &   &   &   & \mbox{\huge 0} &   \\
   &   &   & -\frac{1}{2}\delta_1 &   &   &   &   &   \\
   &   &   &   & -\frac{1}{2}\delta_1 &   &   &   &   \\
   &   &   &   &   & -\frac{1}{2}\delta_2 &   &   &   \\
   & \mbox{\huge 0} &   &   &   &   & -\frac{1}{2}\delta_2 &   &   \\
   &   &   &   &   &   &   & -\frac{1}{2}\delta_3 &   \\
   &   &   &   &   &   &   &   & -\frac{1}{2}\delta_3 \\
\end{array}
\right),
\end{equation}
where parts acting on the occupation probabilities and coherences are separated, and
with $\delta_1 = \kappa_{13} + \kappa_{21}$, $\delta_2 = \kappa_{12} + \kappa_{21} + \kappa_{32}$
and $\delta_3 = \kappa_{12} + \kappa_{13} + \kappa_{32}$.
Producing the cumulants $\tilde{c}_\nu = \{0.64,0.32,0.12,\ldots\}$ and following the same steps as for the atomic system,
we recover the original rates and the additional solution $\{\kappa_{21},\kappa_{12}, \kappa_{32},\kappa_{13}\}=\{3, \frac{1}{3}, \frac{2}{3}, 3\}$.
We thus find two generators that are not trivially related and yield the same counting
statistics for the product molecules. The original rates describe an enzyme with a low enzyme-substrate
binding efficiency whereas the second solution identifies a bottleneck in the substrate-product
conversion (cf.~Fig.~\ref{bio}).

The steady-state occupation probabilities are $p=\{\frac{9}{14},\frac{3}{14}, \frac{1}{7}\}$
and $p=\{\frac{3}{14},\frac{9}{14}, \frac{1}{7}\}$ for the original and additional solution,
respectively, and the classical characteristic polynomial reads
\begin{equation}
	P_\xi^{\rm cl}(x) = x^3 + 7x^2 + 14x - 6(\ee^\xi - 1)\,.
\end{equation}
The sum of the rates $\kappa_{ij}$ is identical to the coefficient $a_{N-1}$ of $P_\xi^{\rm cl}(x)$, which holds
for any stochastic system and provides a convenient consistency check. We see that in
this specific case the counting statistics of the observed product molecules is not
sufficient to determine the rates $\kappa_{ij}$ uniquely. Occupation probabilities
would nevertheless provide the possibility to discriminate between the two sets of rates.


\section{Conclusions}
\label{sfive}

We have addressed the problem of finding the properties of a finite-dimensional
stochastic or open quantum system from the counting statistics of observable
time-discrete events, a scenario encountered in many experimental situations.
The first crucial step toward the solution of this newly posed problem was the
reconstruction the characteristic polynomial of the deformed generator from a
finite number of cumulants by merely solving a linear system. It was moreover
shown that the cumulant generating function can be replaced by characteristic
polynomial, from which all cumulants can be determined recursively.

By exploiting the fact that only a finite number of the cumulants are independent
we have proposed cumulant-based tests for identifying non-classicality, non-Markovianity
and a lower bound for the system dimension. In particular, the non-classicality of a Markovian
system with dimension $N$ can be identified by measuring the first $2N-1$ cumulants.
In contrast to specific state preparations required for similar tests, we only require
the system to be in the steady state as we utilize information that naturally leaks out
of the system.

As the second important step of ICS, we have suggested a direct analytical approach to the reconstruction
of the generator. While perfectly adequate for systems of small dimension this approach requires
the solution of potentially large polynomial systems, currently a limiting factor of ICS. Formulated
in terms of the spectrum $\sigma_\xi$, the reconstruction of the generator falls into the class of
structured inverse eigenvalue problems (SIEP)~\cite{Chu-AN-2002,Kirsch-2011}. In order to apply ICS
to larger systems and to cope with inevitable measurement errors we plan to develop efficient and robust
numerical methods to solve the polynomial system in Eq.~\eref{poly2} or the equivalent SIEP.

The results of ICS also shed light on the established theory of FCS. It was shown that different
generators, i.e., different sets of system parameters, can lead to identical observed counting statistics.
Our reconstruction procedure suggests that these equivalent generators form finite discrete sets. The
problem of non-uniqueness, an essential aspect of counting statistics, has so far not been directly addressed
in the framework of FCS. The characteristic polynomial approach to calculating (zero-frequency)
cumulants according to Eqs.~\eref{cm} offers in addition a potentially useful alternative to the traditional
methods of FCS.

Measuring cumulants of high order is experimentally demanding, but in principal a problem
of acquiring sufficient data to determine the cumulants with the required precision. Cumulants
up to $15$th order can be measured in electronic currents through mesoscopic systems~\cite{Flindt-PNAS-2009}
and high order cumulants have been recently used to characterize $^{87}$Rb spin ensembles prepared in
non-Gaussian states~\cite{Dubost-PRL-2012}. Interesting stochastic systems to which ICS can be
applied are abundant, for example, data transfer in computer networks, traffic problems
and biological processes. We believe that ICS as an analytical tool will contribute to a
better understanding of these systems.



\ack\vspace{-5pt}
MB, LDPC and MBP are supported by the ERC Synergy grant BioQ, the EU
Integrating project SIQS, the EU STREP project PAPETS and the Alexander
von Humboldt Foundation.


\section*{References}

\bibliographystyle{with_titles}

\bibliography{counting}

\begin{thebibliography}{10}

\bibitem{Diedrich-PRL-1987}
F.~Diedrich and H.~Walther.
\newblock Nonclassical radiation of a single stored ion.
\newblock {\em Phys. Rev. Lett.}{ \bf 58}, 203 (1987).

\bibitem{Jelezko-PSS-2006}
F.~Jelezko and J.~Wrachtrup.
\newblock Single defect centres in diamond: a review.
\newblock {\em Phys. Status Solidi A}{ \bf 203}, 3207 (2006).

\bibitem{Gustavsson-PRL-2006}
S.~Gustavsson, R.~Leturcq, B.~Simovi{\v{c}}, R.~Schleser, T.~Ihn, P.~Studerus,
  K.~Ensslin, D.~C. Driscoll, and A.~C. Gossard.
\newblock Counting statistics of single electron transport in a quantum dot.
\newblock {\em Phys. Rev. Lett.}{ \bf 96}, 076605 (2006).

\bibitem{Gustavsson-SSR-2009}
S.~Gustavsson, R.~Leturcq, M.~Studer, I.~Shorubalko, T.~Ihn, K.~Ensslin, D.~C.
  Driscoll, and A.~C. Gossard.
\newblock Electron counting in quantum dots.
\newblock {\em Surf. Sci. Rep.}{ \bf 64}, 191 (2009).

\bibitem{Flindt-PNAS-2009}
C.~Flindt, C.~Fricke, F.~Hohls, T.~Novotn{\`y}, K.~Neto{\v{c}}n{\`y},
  T.~Brandes, and R.~J. Haug.
\newblock Universal oscillations in counting statistics.
\newblock {\em Proc. Natl. Acad. Sci.}{ \bf 106}, 10116 (2009).

\bibitem{Kolomeisky-ARPC-2007}
A.~B. Kolomeisky and M.~E. Fisher.
\newblock Molecular motors: a theorist's perspective.
\newblock {\em Annu. Rev. Phys. Chem.}{ \bf 58}, 675 (2007).

\bibitem{English-NCB-2005}
B.~P. English, W.~Min, A.~M. van Oijen, K.~T. Lee, G.~Luo, H.~Sun, B.~J.
  Cherayil, S.C. Kou, and X.~S. Xie.
\newblock Ever-fluctuating single enzyme molecules: {M}ichaelis-{M}enten
  equation revisited.
\newblock {\em Nat. Chem. Biol.}{ \bf 2}, 87 (2005).

\bibitem{Moffitt-ME-2010}
J.~R. Moffitt, Y.~R. Chemla, and C.s Bustamante.
\newblock Methods in statistical kinetics.
\newblock {\em Methods Enzymol.}{ \bf 475}, 221 (2010).

\bibitem{Moffitt-FEBS-2013}
J.~R. Moffitt and C.~Bustamante.
\newblock Extracting signal from noise: kinetic mechanisms from a
  {M}ichaelis--{M}enten-like expression for enzymatic fluctuations.
\newblock {\em FEBS Journal}{ \bf } (2013).

\bibitem{Kirsch-2011}
A.~Kirsch.
\newblock {\em An introduction to the mathematical theory of inverse problems}.
\newblock Springer,  (2011).

\bibitem{Vogel-PRA-1989}
K.~Vogel and H.~Risken.
\newblock Determination of quasiprobability distributions in terms of
  probability distributions for the rotated quadrature phase.
\newblock {\em Phys. Rev. A}{ \bf 40}, 2847 (1989).

\bibitem{Paris-2004}
M.~G.~A. {Paris} and J.~{{\v R}eh{\'a}{\v c}ek}, editors.
\newblock {\em {Quantum State Estimation}}, volume 649 of {\em Lecture Notes in
  Physics, Springer},  (2004).

\bibitem{Baumgratz-PRL-2013}
T.~Baumgratz, D.~Gross, M.~Cramer, and M.~B. Plenio.
\newblock Scalable reconstruction of density matrices.
\newblock {\em Phys. Rev. Lett.}{ \bf 111}, 020401 (2013).

\bibitem{Baumgratz-AX-2013}
T.~Baumgratz, A.~N{\"u}{\ss}eler, M.~Cramer, and M.~B. Plenio.
\newblock A scalable maximum likelihood method for quantum state tomography.
\newblock {\em Preprint arXiv:1308.2395}{ \bf } (2013).
\newblock Accepted for publication in NJP.

\bibitem{Leggett-PRL-1985}
A.~J. Leggett and A.~Garg.
\newblock Quantum mechanics versus macroscopic realism: {I}s the flux there
  when nobody looks?
\newblock {\em Phys. Rev. Lett.}{ \bf 54}, 857 (1985).

\bibitem{Emary-AX-2013}
C.~Emary, N.~Lambert, and F.~Nori.
\newblock {L}eggett-{G}arg inequalities.
\newblock {\em Preprint arXiv:1304.5133}{ \bf } (2013).

\bibitem{Waldherr-PRL-2011}
G.~Waldherr, P.~Neumann, S.~F. Huelga, F.~Jelezko, and J.~Wrachtrup.
\newblock Violation of a temporal {B}ell inequality for single spins in a
  diamond defect center.
\newblock {\em Phys. Rev. Lett.}{ \bf 107}, 090401 (2011).

\bibitem{Lambert-PRL-2010}
N.~Lambert, C.~Emary, Y.-N. Chen, and F.~Nori.
\newblock Distinguishing quantum and classical transport through
  nanostructures.
\newblock {\em Phys. Rev. Lett.}{ \bf 105}, 176801 (2010).

\bibitem{Bednorz-PRL-2010}
A.~Bednorz and W.~Belzig.
\newblock Quasiprobabilistic interpretation of weak measurements in mesoscopic
  junctions.
\newblock {\em Phys. Rev. Lett.}{ \bf 105}, 106803 (2010).

\bibitem{Bednorz-PRB-2011}
A.~Bednorz and W.~Belzig.
\newblock Proposal for a cumulant-based {B}ell test for mesoscopic junctions.
\newblock {\em Phys. Rev. B}{ \bf 83}, 125304 (2011).

\bibitem{Wolf-PRL-2008}
M.~M. Wolf, J.~Eisert, T.~S. Cubitt, and J.~I. Cirac.
\newblock Assessing non-{M}arkovian quantum dynamics.
\newblock {\em Phys. Rev. Lett.}{ \bf 101}, 150402 (2008).

\bibitem{Breuer-PRL-2009}
H.-P. Breuer, E.-M. Laine, and J.~Piilo.
\newblock Measure for the degree of non-{M}arkovian behavior of quantum
  processes in open systems.
\newblock {\em Phys. Rev. Lett.}{ \bf 103}, 210401 (2009).

\bibitem{Rivas-PRL-2010}
{\'A}.~Rivas, S.~F. Huelga, and M.~B. Plenio.
\newblock Entanglement and non-{M}arkovianity of quantum evolutions.
\newblock {\em Phys. Rev. Lett.}{ \bf 105}, 050403 (2010).

\bibitem{Brunner-PRL-2008}
N.~Brunner, S.~Pironio, A.~Acin, N.~Gisin, A.~A. M{\'e}thot, and V.~Scarani.
\newblock Testing the dimension of {H}ilbert spaces.
\newblock {\em Phys. Rev. Lett.}{ \bf 100}, 210503 (2008).

\bibitem{Hendrych-NP-2012}
M.~Hendrych, R.~Gallego, M.~Mi{\v{c}}uda, N.~Brunner, A.~Ac{\'\i}n, and J.~P.
  Torres.
\newblock Experimental estimation of the dimension of classical and quantum
  systems.
\newblock {\em Nat. Physics}{ \bf 8}, 588 (2012).

\bibitem{Levitov-JMP-1996}
L.~S. Levitov, H.-W. Lee, and G.~B. Lesovik.
\newblock Electron counting statistics and coherent states of electric current.
\newblock {\em J. Math. Phys.}{ \bf 37}, 4845 (1996).

\bibitem{Belzig-PRL-2001}
W.~Belzig and Y.~V. Nazarov.
\newblock Full counting statistics of electron transfer between
  superconductors.
\newblock {\em Phys. Rev. Lett.}{ \bf 87}, 197006 (2001).

\bibitem{Bagrets-PRB-2003}
D.~A. Bagrets and Y.~V. Nazarov.
\newblock Full counting statistics of charge transfer in {C}oulomb blockade
  systems.
\newblock {\em Phys. Rev. B}{ \bf 67}, 085316 (2003).

\bibitem{Sanchez-PRL-2007}
R.~S{\'a}nchez, G.~Platero, and T.~Brandes.
\newblock Resonance fluorescence in transport through quantum dots: noise
  properties.
\newblock {\em Phys. Rev. Lett.}{ \bf 98}, 146805 (2007).

\bibitem{Flindt-PRL-2008}
C.~Flindt, T.~Novotn{\`y}, A.~Braggio, M.~Sassetti, and A.-P. Jauho.
\newblock Counting statistics of non-{M}arkovian quantum stochastic processes.
\newblock {\em Phys. Rev. Lett.}{ \bf 100}, 150601 (2008).

\bibitem{Flindt-PRB-2010}
C.~Flindt, T.~Novotn{\`y}, A.~Braggio, and A.-P. Jauho.
\newblock Counting statistics of transport through {C}oulomb blockade
  nanostructures: high-order cumulants and non-{M}arkovian effects.
\newblock {\em Phys. Rev. B}{ \bf 82}, 155407 (2010).

\bibitem{Nazarov-2009}
Y.~V. Nazarov and Y.~M. Blanter.
\newblock {\em Quantum transport: introduction to nanoscience}.
\newblock Cambridge University Press,  (2009).

\bibitem{Derrida-PRL-1998}
B.~Derrida and J.~L. Lebowitz.
\newblock Exact large deviation function in the asymmetric exclusion process.
\newblock {\em Phys. Rev. Lett.}{ \bf 80}, 209 (1998).

\bibitem{Lebowitz-JSP-1999}
J.~L. Lebowitz and H.~Spohn.
\newblock A {G}allavotti--{C}ohen-type symmetry in the large deviation
  functional for stochastic dynamics.
\newblock {\em J. Stat. Phys.}{ \bf 95}, 333 (1999).

\bibitem{Derrida-JSM-2007}
B.~Derrida.
\newblock Non-equilibrium steady states: fluctuations and large deviations of
  the density and of the current.
\newblock {\em J. Stat. Mech.: Theor. Exp.}{ \bf 2007}, P07023 (2007).

\bibitem{Touchette-PR-2009}
H.~Touchette.
\newblock The large deviation approach to statistical mechanics.
\newblock {\em Phys. Rep.}{ \bf 478}, 1 (2009).

\bibitem{Reulet-PRL-2003}
B.~Reulet, J.~Senzier, and D.~E. Prober.
\newblock Environmental effects in the third moment of voltage fluctuations in
  a tunnel junction.
\newblock {\em Phys. Rev. Lett.}{ \bf 91}, 196601 (2003).

\bibitem{Bomze-PRL-2005}
Yu. Bomze, G.~Gershon, D.~Shovkun, S.~Levitov, L.~\, and M.~Reznikov.
\newblock Measurement of counting statistics of electron transport in a tunnel
  junction.
\newblock {\em Phys. Rev. Lett.}{ \bf 95}, 176601 (2005).

\bibitem{Kampen-1992}
N.~G. Van~Kampen.
\newblock {\em Stochastic processes in physics and chemistry}.
\newblock Elsevier,  (2007).

\bibitem{Breuer-2002}
H.~P. Breuer and F.~Petruccione.
\newblock {\em The theory of open quantum systems}.
\newblock Oxford University Press,  (2002).

\bibitem{Rivas-2011}
{\'A}.~Rivas and S.~F. Huelga.
\newblock {\em Open quantum systems. {A}n introduction}.
\newblock Springer,  (2011).

\bibitem{Monras-AX-2010}
A.~Monras, A.~Beige, and K.~Wiesner.
\newblock Hidden quantum markov models and non-adaptive read-out of many-body
  states.
\newblock {\em Preprint arXiv:1002.2337}{ \bf } (2010).

\bibitem{Zoller-PRA-1987}
P.~Zoller, M.~Marte, and D.~F. Walls.
\newblock Quantum jumps in atomic systems.
\newblock {\em Phys. Rev. A}{ \bf 35}, 198 (1987).

\bibitem{Raghunathan-PIAS-1981}
P.~Raghunathan.
\newblock The characteristic polynomial approach to the solution of quantum
  chemical perturbation problems.
\newblock {\em Proc. Indian Acad. Sci. (Chem. Sci.)}{ \bf 90}, 467 (1981).

\bibitem{Macduffee-2004}
C.~C. MacDuffee.
\newblock {\em The theory of matrices}.
\newblock Courier Dover Publications,  (2004).

\bibitem{Horn-2012}
R.~A. Horn and C.~R. Johnson.
\newblock {\em Matrix analysis}.
\newblock Cambridge University Press,  (2012).

\bibitem{Cox-2007}
D.~A. Cox.
\newblock {\em Ideals, varieties, and algorithms: an introduction to
  computational algebraic geometry and commutative algebra}.
\newblock Springer,  (2007).

\bibitem{Reilly-2010}
N.~R. Reilly.
\newblock {\em Introduction to applied algebraic systems}.
\newblock Oxford University Press,  (2010).

\bibitem{Contreras-TBP-2013}
L.~D. Contreras-Pulido et~\emph{al.}
\newblock (to be published).

\bibitem{Fleischhauer-RMP-2005}
M.~Fleischhauer, A.~Imamoglu, and J.~P. Marangos.
\newblock Electromagnetically induced transparency: optics in coherent media.
\newblock {\em Rev. Mod. Phys.}{ \bf 77}, 633 (2005).

\bibitem{Qian-BC-2002}
H.~Qian and E.~L. Elson.
\newblock Single-molecule enzymology: stochastic {M}ichaelis-{M}enten kinetics.
\newblock {\em Biophysical chemistry}{ \bf 101}, 565--576 (2002).

\bibitem{Chu-AN-2002}
M.~T. Chu and G.~H. Golub.
\newblock Structured inverse eigenvalue problems.
\newblock {\em Acta Numerica}{ \bf 11}, 1 (2002).

\bibitem{Dubost-PRL-2012}
B.~Dubost, M.~Koschorreck, M.~Napolitano, N.~Behbood, R.~J. Sewell, and M.~W.
  Mitchell.
\newblock Efficient quantification of non-{G}aussian spin distributions.
\newblock {\em Phys. Rev. Lett.}{ \bf 108}, 183602 (2012).

\end{thebibliography}

\end{document}